\documentstyle[aps,multicol,epsfig]{revtex}
\begin{document}
\draft
\title{Semiflexible Polymer in a Strip}
\author{J\"{u}rgen F. Stilck}
\address{
Instituto de F\'{\i}sica\\
Universidade Federal Fluminense\\
Av. Litor\^anea, s/n\\
24210-340 - Niter\'oi - RJ\\
Brazil}
\date{\today}
\maketitle
\begin{abstract}
We study the thermodynamic properties of a semiflexible polymer
confined inside strips of widths $L \leq 9$ defined on a square lattice.
The polymer is modeled as a self-avoiding walk and a short-range
interaction between the monomers and the walls is included through an
energy $\epsilon$ associated to each monomer placed on one of the
walls. Also, an energy $ \epsilon_b$ is associated to each elementary
bend of the walk. The free energy of the model is obtained exactly
through a transfer matrix formalism. The profile of monomer density
and the force on the walls are obtained. We notice that as $\epsilon_b$ is
decreased, the range of values for $\epsilon$ for which the density profile is
neither convex nor concave increases, and for sufficiently attracting
walls ($\epsilon<0$) we find that in general the attractive
force is maximum for $\epsilon_b<0$, that is, for situations where
the bends are favored.
\end{abstract}
\pacs{05.50.+q, 61.41.+e}

\begin{multicols}{2}

\section{Introduction}
\label{I}

Polymers are often modeled as self- and mutually avoiding walks
placed on a lattice, and much has been learned about their
thermodynamic properties through such models \cite{f53,dg79}.
Grand-canonical models of this kind, where the number of monomers
incorporated into the chain is allowed to fluctuate controlled by an
activity $z=\exp(\mu/k_BT)$, display a phase transition at some value
of the activity (for infinite chains, that is, in the polymer limit).
This transition is discontinuous in the one dimensional case $d=1$
\cite{pw83}, and continuous for $d>2$. Rather precise estimates of
the critical value of $z$ were found in two dimensions through
transfer matrix calculations \cite{d81} and series expansions
\cite{cg96}. Also, exact values for the critical exponents are
available in this case \cite{n82}. 

More recently, properties of such models on lattices limited by walls
have attracted much interest \cite{dl93}, following studies of
magnetic models in the same situation \cite{bi83}. The short range
interaction between the wall and the polymer may be introduced
associating a Boltzmann factor $\omega=\exp(-\epsilon/k_BT)$ to each
monomer placed on the wall, so that $\epsilon<0$ corresponds to
attracting walls while repulsive walls are described by $\epsilon>0$.
The grand-canonical partition function for a model of a single chain
is
\begin{equation}
Y(x,\omega)=\sum z^N \omega^{N_w},
\label{e1}
\end{equation}
where $N$ is the number of monomers in the chain, the sum is over all
configurations of the chain with the initial monomer placed on the
wall, and $N_w \leq N$ stands for the number of monomers located on
the wall. Such a model shows interesting features, and even in the
limit where the self avoidance constraint is neglected (the so-called
{\em ideal} chains) one finds that for sufficiently large values of
$\omega>\omega_0$ the surface polymerization transition will occur at a lower
value of the activity $z$ than the one in the bulk \cite{r65}. The
point $(z_0,\omega_0)$ in the phase diagram where the bulk (also called
ordinary) transition line meets the surface transition line is called
the adsorption transition point. In two dimensions, such models have
been studied through transfer matrix calculations \cite{gb89} and
series expansions \cite{zld90}.  Additional walls may be added,
confining the polymer inside a strip, slab or pore \cite{dl93}, and
the force applied on the walls in such situations is of interest even
from the point of view of applications of polymers as adhesives
\cite{dr71}. The model of ideal chains confined in a slab has been
studied in the past \cite{dr71,sm98}, and it was found that the force
on the walls is attractive if $\omega$ exceeds the adsorption
value $\omega_0$. In the case of self-avoiding chains confined in a
strip on the square lattice, transfer matrix calculations show that
attractive forces appear for $\omega$ below $\omega_0$ \cite{sm98}.
Also, the profile of the monomer density inside the strip was
obtained \cite{s97}, and, unlike what happens for ideal chains, for a
self-avoiding chain profiles which are neither convex nor concave are
found for a range of values of $\omega$. Finally, similar techniques
have also been applied to shed light on the scaling behavior of such
models for strips with periodic boundary conditions (cylinders) as
the width of the strips becomes large \cite{bg99}.

Another generalization of the original polymer model is to introduce
an energy associated to the operation of bending the chain. On hypercubic
lattices,
the elementary bends will always be at right angles, and an energy
$\epsilon_b$ may be associated to each of them.  This semiflexible
polymer problem (also called persistent or biased walks), has been
studied some time ago \cite{ts75,mn91,cfs92}. Recently, this model has
attracted renewed interest , since it may 
describe some relevant aspects in the protein folding problem
\cite{dgh96}. The thermodynamic properties of the model have been
studied on the Bethe lattice \cite{bs93} and, equivalently, in the
Bethe approximation \cite{lmp98}. The end-to-end distance of
semi-flexible chains on Bethe and Husimi lattices was obtained
\cite{sca00}. 
 In this paper we study the thermodynamic behavior of a
semiflexible polymer confined inside a strip. The partition function
of the model may be written as
\begin{equation}
Y(x,\omega)=\sum z^N \omega^{N_w} \omega_b^{N_b},
\label{e2}
\end{equation}
where $N_b$ is the number of elementary bends in the configuration
and $\omega_b=\exp(-\epsilon_b/k_BT)$ is the Boltzmann factor
associated to each elementary bend and the sum is over all configurations of
the chain. We define a transfer matrix for
the model and obtain the grand-canonical partition function in the
thermodynamic limit, determined by the largest eigenvalue of this
matrix. In order to be able to obtain the distribution of monomers in
the transverse section of the strip, we define a position dependent
activity for the monomers. Also, the force applied by the polymer on
the walls is calculated, as a function of the width of the strip,
$\omega$ and $\omega_b$. All thermodynamic properties are calculated
at the polymerization transition, that is, at the value of the
activity of a monomer for which the number of monomers incorporated
into the polymer diverges. 

In section \ref{M} the model is defined in detail and its solution is
presented. The results we find for the thermodynamic behavior of the
model are shown in section \ref{RC} as well as final conclusions and
discussions.

\section{Definition of the model and its solution}
\label{M}

\begin{figure}
\centerline{\epsfig{file=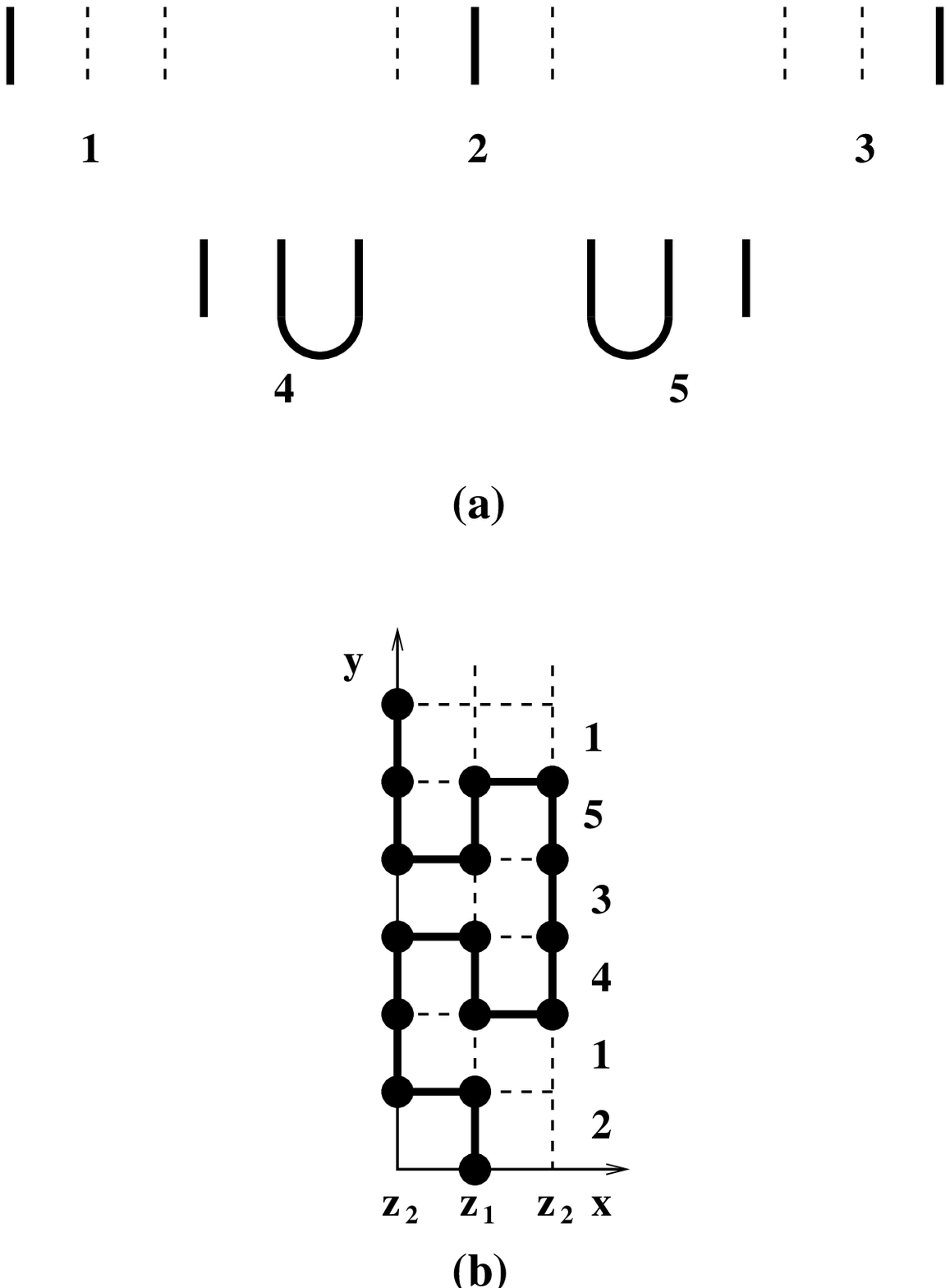,width=8cm,angle=0} }
\caption{a) The five connectivity configurations for $m=2$. b) Portion
of the chain, with the number of each connectivity configuration
indicated}
\label{conf}
\end{figure}

The self avoiding chain is constrained inside a strip of width $m$
defined on the square lattice in the $(x,y)$ plane, so that $0 \le x
\le m$. The chain runs through the whole strip, from $y \to -\infty$
to $y \to +\infty$. We may define a transfer matrix for the problem
following a prescription proposed by Derrida, in a way to take into
account the self-avoidance constraint exactly \cite{d81}. The
connectivity properties of all vertical bonds of the chain arriving at
a line $y_0$ from below are specified through the indication of
\begin{enumerate}

\item The (unique) bond connected to the initial monomer of the chain
(placed in $y \to -\infty$) through a path lying entirely below the
line $y_0$ (passing though sites with $y<y_0$);

\item The pairs of bonds connected to each other through a path lying
entirely below the line $y_0$.

\end{enumerate}
In figure \ref{conf} the five configurations for the case $m=2$ are
depicted, and a portion of the chain placed inside the strip is shown
with each configuration indicated. Configurations 1 and 3, as well as
4 and 5, are related to each other by reflection symmetry. We define
column dependent activities $z_i$, $i=1, 2, ...n_a(m)$ according to
this reflection symmetry, as indicated also in figure \ref{conf}. The
number of activities $n_a(m)$ is equal to $m/2+1$ for even values of
$m$ and $(m+1)/2$ for odd values of $m$.

\begin{figure}
\centerline{\epsfig{file=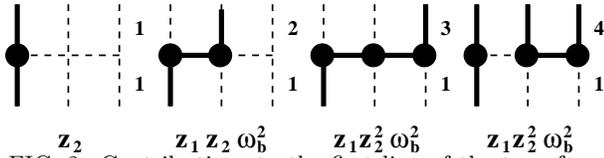,width=8cm,angle=0} }
\caption{Contributions to the first line of the transfer matrix for
the case $m=2$. The connectivity configurations are indicated
following the enumeration adopted in figure \ref{conf}}
\label{tmc}
\end{figure}

For a fixed connectivity configuration of a set of $m+1$ vertical
bonds arriving at the line $y_0$, the possible configurations of the
set of vertical bonds arriving at $y_0+1$ may be obtained, as well as
the contribution to the partition function from the sites comprised
between both sets of vertical bonds. This contribution will be given
by 
\[
\omega_b^{N_{b,y_0}}\prod_{i=1}^{n_a(m)} z_i^{N_{i,y_0}}
\]
where $N_{i,y_0}$ is the number of monomers with activity $z_i$ in line
$y_0$ and $N_{b,y_0}$ is the number of elementary bends in this line. These
contributions, shown for a particular example in figure \ref{tmc},
define a line of the transfer matrix. For $m=2$ the transfer matrix
will be given by
\begin{equation}
{\bf T}=
\left(
\begin{array}{ccccc}
z_2&z_1z_2\omega_b^2&z_1z_2^2\omega_b^2&z_1z_2^2\omega_b^2&0 \\
z_1z_2\omega_b^2&z_1&z_1z_2\omega_b^2&0&0 \\
z_1z_2^2\omega_b^2&z_1z_2\omega_b^2&z_2&0&z_1z_2^2\omega_b^2\\
0&0&z_1z_2^2\omega_b^2&z_1z_2^2&0\\
z_1z_2^2\omega_b^2&0&0&0&z_1z_2^2 \\
\end{array}
\right).
\label{e3}
\end{equation}

The grand canonical partition function of the model,considering
periodic boundary conditions in the $y$ direction, will be given by
\begin{equation}
\Xi=\mathrm(Tr){\bf T}^{N_y},
\label{e4}
\end{equation}
where $N_y$ is the total length of the strip in the $y$ direction, so
that the total number of sites is given by $N_s=N_y(m+1)$. The number
of monomers with activity $z_i$ is
\begin{equation}
N_i=\frac{z_i}{\Xi}\frac{\partial \Xi}{\partial z_i},
\label{e5}
\end{equation}
and the total number of monomers in the chain will be
\begin{equation}
N=\sum_{i=1}^{n_a(m)} N_i.
\label{e6}
\end{equation}
The fraction of monomers placed at column $x$ is
\begin{equation}
\rho(x)=\frac{N_{x+1}}{(2-\delta_{x,m/2})N},
\label{e7}
\end{equation}
where $x=0,1,\ldots,n_a(m)-1$ and $\rho(x)=\rho(m-x)$. The Kronecker
delta in the denominator contributes only for even values of $m$.

In the thermodynamic limit $N_y \to \infty$ the partition function
\ref{e4} is dominated by the largest eigenvalue of the transfer
matrix $\lambda_1$, so that $\Xi \sim \lambda_1^{N_y}$ and in this
limit 
\begin{equation}
N_i=N_y \frac{z_i}{\lambda_1}\frac{\partial \lambda_1}{\partial z_i}.
\label{e8}
\end{equation}
The first order polymerization transition in the finite strip will
take place when the thermodynamic potential 
\begin{equation}
\psi=-k_BT \ln(\Xi)
\label{e9}
\end{equation}
is equal to the one for the empty lattice $\psi_0=0$. Thus, since in
the thermodynamic limit we have
\begin{equation}
\psi/N_s=-k_BT(m+1)\ln(\lambda_1)
\label{e10}
\end{equation}
the polymerized phase will coexist with the non-polymerized phase for
$\lambda_1=1$. Therefore, all thermodynamic quantities below will be
calculated for $z_i=z, i=1,\ldots,n_a(m)-1$ and $z(n_a(m))=\omega z$,
where the activity $z$ is then fixed at the coexistence value $z_c$
so that, for a given value of $\omega_b$, we have $\lambda_1=1$.

Finally, the force applied on the walls is given by
\begin{equation}
F=\frac{1}{a}\left( \frac{\partial \Xi}{\partial
m}\right)_{z,\omega,\omega_b}, 
\label{e11}
\end{equation}
where $a$ is the lattice parameter and positive values correspond to
attractive forces. An adimensional force per monomer at the
coexistence may be then defined as
\begin{equation}
f=\frac{Fa}{k_BTN}=\frac{-\left( \frac{\partial \Xi}{\partial m}
\right)_{z,\omega,\omega_b} }{z \left( \frac{\partial \Xi}{\partial
z}\right)_{m,\omega,\omega_b}}=\frac{1}{z_c}\left(\frac{\partial
z_c}{\partial m}\right)_{\omega,\omega_b}.
\label{e12}
\end{equation}
Since our results correspond to integer values of $m$, the force $f$
was estimated making the discrete approximation
\begin{eqnarray}
f(m+1/2,\omega,\omega_b) &\approx& \frac{2}{z_c(m+1,\omega,\omega_b)
+ z_c(m,\omega,\omega_b)} \nonumber \\
& & \times [z_c(m+1,\omega,\omega_b)-z_c(m,\omega,\omega_b)].
\label{e13}
\end{eqnarray}

A simple calculation may be performed in the limit of rigid rods $\omega_b=0$,
where bends are not allowed. In this limit the transfer matrix is diagonal
of size $(m+1) \times (m+1)$. For $\omega>1$ all monomers are on the walls,
for $\omega=1$ they are uniformly distributed, whereas for $\omega<1$ an
uniform distribution in the sites away from the walls is found. As expected,
in this limit the force on the walls vanishes. In the numerical results below
we consider $\omega_b \geq 1$.

\section{Numerical results and conclusion}
\label{RC}

The transfer matrices for the model were obtained exactly for strips of widths
ranging from 3 to 9. After using the reflection symmetry, the sizes of the
matrices are, respectively, equal to 6, 16, 38, 100, 256, 681, and 1805. We
obtained the density profile at the coexistence condition for values of
$\omega$ and $\omega_b$ mostly between 1 and 3.

\begin{figure}
\centerline{\epsfig{file=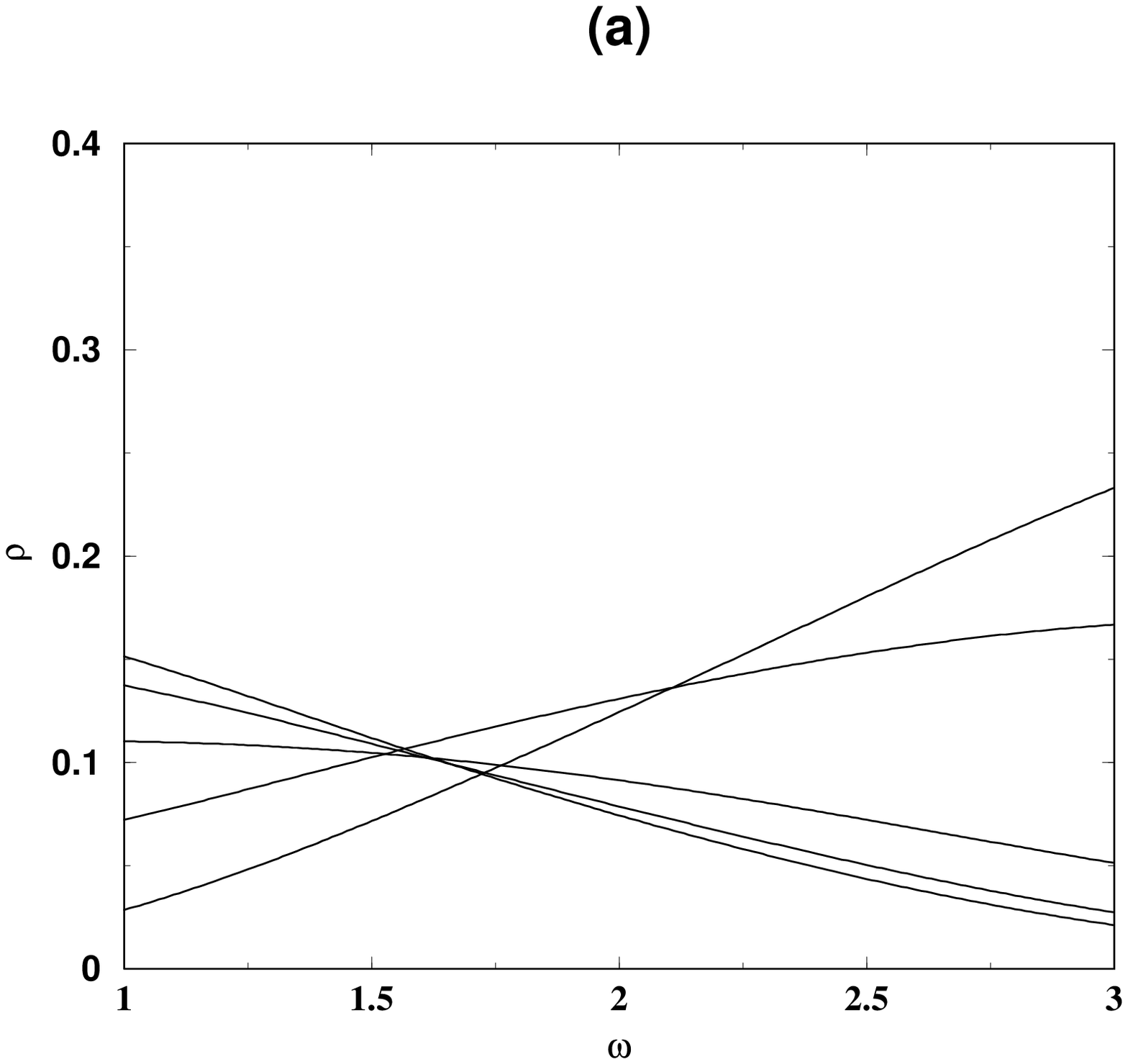,width=8cm,angle=0} }
\centerline{\epsfig{file=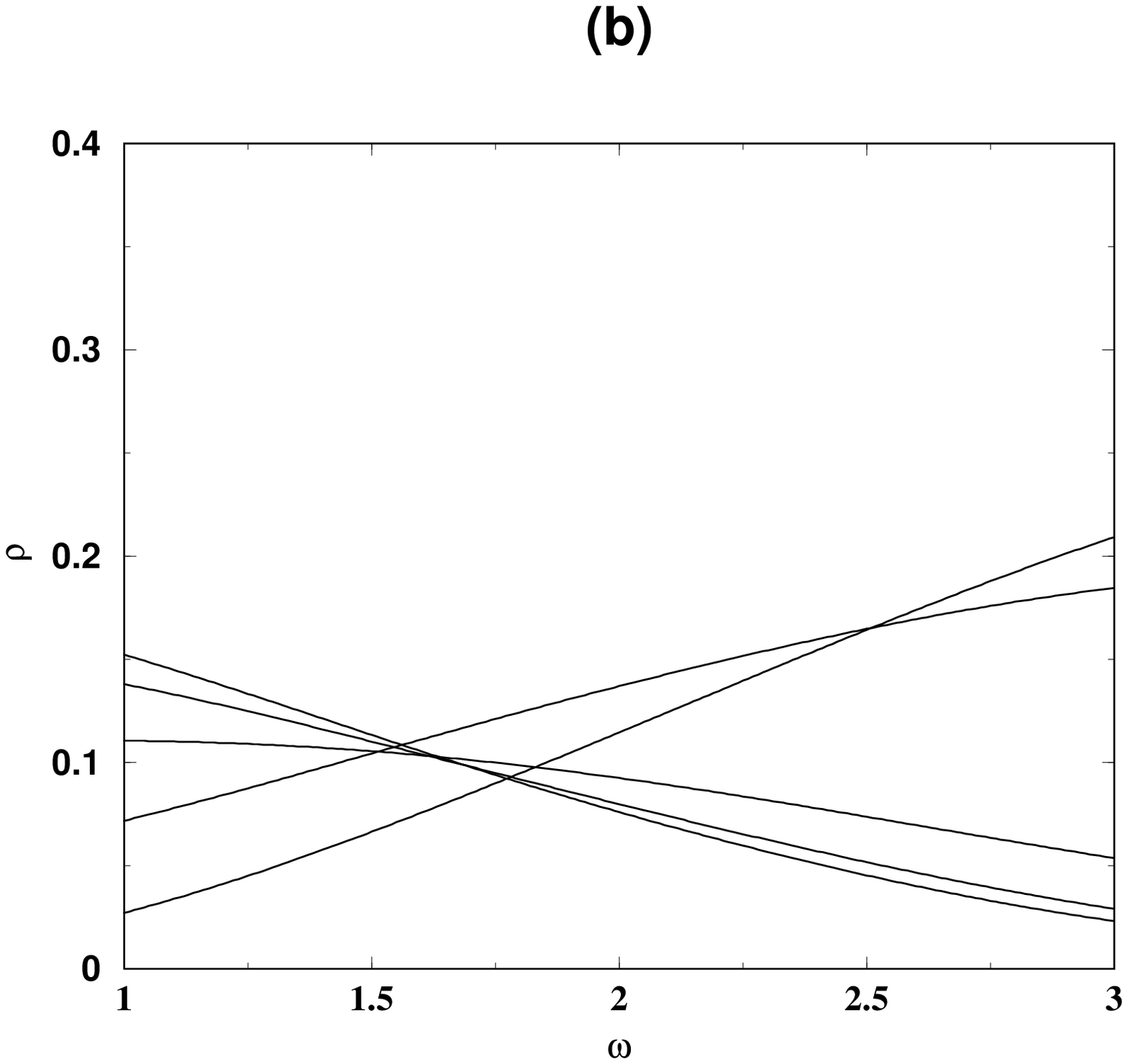,width=8cm,angle=0}}
\caption{Density of monomers as function of $\omega$ for a strip with
$m=9$. (a) corresponds to $\omega_b=2$ and (b) to $\omega_b=3$. Curves
are for different values of $x$ and $\rho(x)=\rho(m-x)$. At $\omega=1$
the highest density corresponds to the center of the strip ($x=4$ and $x=5$)
and the density decreases monotonically outwards, being lowest at the wall
($x=0$ and $x=9$). At $\omega=3$ the density profile is convex in both cases,
with the maximum density located on the walls.} 
\label{dts}
\end{figure}

For neutral walls ($\omega=1$) the density profile is always concave, with a
higher density in the center. This may be understood since the region away
from the walls is favored entropically. As $\omega$ becomes larger, monomers
on the walls are energetically favored, so for sufficiently large values of
$\omega$ a convex density profile is expected. For ideal flexible chains
($\omega_b=1$), at the 
adsorption value $\omega=4/3$ the density profile is flat for all sites
which are not on the walls \cite{dr71,s97}. Such a flat transition profile is
not observed for self-avoiding chains, where convex profiles are separated
from concave ones by an interval of values of $\omega$, located well below the
adsorption transition value, where the profile is
neither convex nor concave. This interval is quite narrow for flexible chains
\cite{s97}. In figure \ref{dts} the densities are plotted as functions of
$\omega$ for two values of $\omega_b$. It is clear that as $\omega_b$ is
increased, favoring bends, the interval of values of $\omega$ with a profile
without well defined convexity grows. 

\begin{figure}
\centerline{\epsfig{file=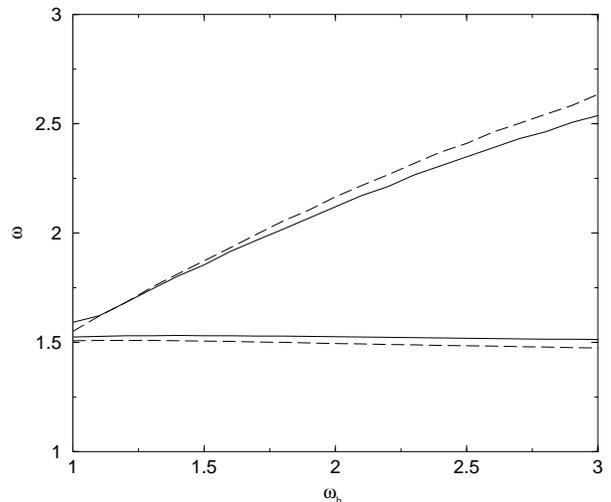,width=8cm,angle=0} }
\caption{Boltzmann weights below which the density is concave (lower curves)
and above which it is convex (upper curves) for $m=6$ (dashed curves) and
$m=9$ (full curves).} 
\label{wc}
\end{figure}

The values of $\omega$ below which the density profile is concave
($\omega_1$)and those 
above which the profile is convex ($\omega_2$) are plotted in figure \ref{wc}
as functions 
of $\omega_b$ for two values of $m$. As is apparent the range with no defined
convexity grows with $\omega_b$ in a nearly linear way. As a general rule, we
found that as $\omega_b$ is increased from 1, the first pair of densities to
cross, destroying concavity, is always correspondent to the columns which are
first and second neighbors to the walls. Also, the last crossing, which turns
the profile convex, is between densities corresponding to the columns at the
wall and the neighbor columns. In the limit $\omega_b \to \infty$ we found
that $\omega_2 \to \infty$.

\begin{figure}
\centerline{\epsfig{file=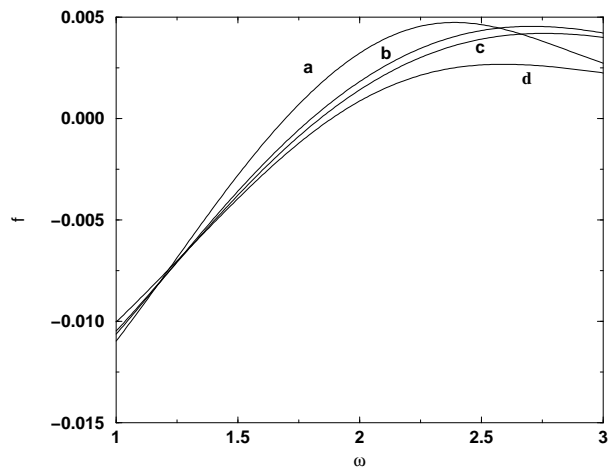,width=8cm,angle=0} }
\caption{Force per monomer on the walls as function of $\omega$ for (a)
$\omega_b=1$, (b) $\omega_b=2$, (c) $\omega_b=3$, and (d) $\omega_b \to
\infty$. Results shown are for $m=8.5$.} 
\label{t}
\end{figure}

Finally, we calculated the force per monomer on the walls for pair of widths
$m$, $m+1$, using expression \ref{e13}. As already found for flexible
self-avoiding chains \cite{sm98}, attractive forces appear for sufficiently
large 
values of $\omega$. Figure \ref{t} shows results for the force as a function
of $\omega$ for some values of $\omega_b$. The origin of attractive forces in
the system are portions of the chain limited by monomers adsorbed on opposite
walls. Thus, as expected, the curves $f \times \omega$ display a maximum,
because the force vanishes as $\omega \to \infty$, since in this limit such
``bridge'' segments are absent.

\begin{figure}
\centerline{\epsfig{file=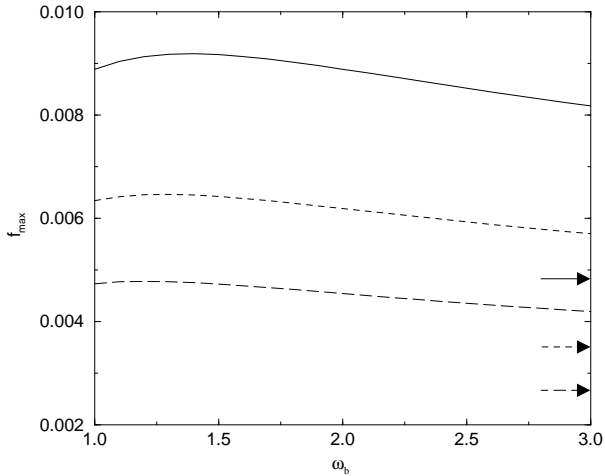,width=8cm,angle=0} }
\caption{Maximum force per monomer on the walls for $m=6.5$ (full line),
$m=7.5$ 
(dotted line), and $m=8.5$ (dashed line) as functions of $\omega_b$. Arrows
indicate the maximum force for $\omega_b \to \infty$.} 
\label{tmax}
\end{figure}

In figure \ref{tmax} the maximum force per monomer (with respect to $\omega$)
is plotted as a function of $\omega_b$. It is apparent that the maximum of
these curves is located at $\omega_b>1$, and thus we conclude that polymers for
which bends are somewhat favored give rise, in general, to larger attractive
forces than flexible ones. As $\omega_b$ is increased, the maximum attractive
force occurs at higher values of $\omega$

The force per monomer as a function of the width of
the strip $m$, for $\omega$ above a threshold, shows a stable equilibrium
point at very low separation $m_1$ and an unstable equilibrium
point at a larger separation $m_2$ \cite{sm98}. The force is attractive in the
interval $m_1<x<m_2$.

The stable equilibrium point $m_1(\omega,\omega_b)$ is found
in the range $0.5 \leq m_1 \leq 1.5$, showing little variation as a
function of the Boltzmann factors, as long as $\omega$ is larger than
the threshold value. The unstable equilibrium point
$m_2(\omega,\omega_b)$ shows a rather strong monotonic dependence on both
variables, being an increasing function of $\omega$ and a decreasing
function of  $\omega_b$.

In conclusion, we may summarize the behavior of semiflexible chains
confined inside strips observing that as the presence of elementary bends
in the chains is favored, the range of values of $\omega$ for which the
density profile has no defined convexity grows. Also, the largest attractive
forces are found for $\omega_b>1$, that is, when the bending of the chains is
favored. It should be noticed that the nonmonotonic behavior observed
for the tension as a function of the width of the strip, with two equilibrium
points, happens in a regime where the finite size scaling behavior has not yet
been reached \cite{sm98}. 

\acknowledgements

Partial financial support by the brazilian agencies CNPq and FAPERJ is
gratefully acknowledged.

\end{multicols}
\end{document}